\documentclass[useAMS,usenatbib]{mn2e}
\usepackage{graphicx}

\def\spose#1{\hbox to 0pt{#1\hss}}
\newcommand\lsim{\mathrel{\spose{\lower 3.0pt\hbox{$\mathchar"218$}}
     \raise 2.0pt\hbox{$\mathchar"13C$}}}
\newcommand\gsim{\mathrel{\spose{\lower 3.0pt\hbox{$\mathchar"218$}}
     \raise 2.0pt\hbox{$\mathchar"13E$}}}
\newcommand\msun{{\rm \,M_\odot}}

\title[Low-End Mass Function of the Arches Cluster]{Low-End Mass Function of the Arches Cluster}
\author[Jihye Shin and Sungsoo S. Kim]{Jihye
Shin$^{1,2}$\thanks{Email: jhshin.jhshin@gmail.com}
and Sungsoo S. Kim$^{2,3}$\thanks{Corresponding author}\\
$^{1}$Kavli Institute for Astronomy and Astrophysics at Peking
University, Yi He Yuan Lu 5, Hai Dian District, Beijing 100871, P.R. China\\
$^{2}$School of Space Research, Kyung Hee University, Yongin, Kyungki
446-701, Republic of Korea\\
$^{3}$Department of Astronomy \& Space Science, Kyung Hee University,
Yongin, Kyungki 446-701, Republic of Korea}
\begin{document}

\date{Accepted 2014 November 12. Received 2014 November 11; in original form 2014 October 5}

\pagerange{\pageref{firstpage}--\pageref{lastpage}} \pubyear{2014}

\maketitle

\label{firstpage}

\begin{abstract}
The initial mass function (IMF) of the Arches cluster, which was formed a few
million years ago in the harsh environment of the Galactic center (GC),
has long been a target of interest to those who study the GC and the theory
of star formation.  The distinct star-forming conditions in the GC might
have caused the cluster to have a shallower slope or an elevated
lower mass cutoff in its IMF.  But its mass function has been revealed
only down to 1-2$\msun$ (the lower limit of resolved stars), and the
low-end mass function of the Arches is still unknown.
To estimate the unresolved part of the Arches mass function,
we have devised a novel photometric method that involves the histogram of
pixel intensities in the observed image, which contains information
on the unresolved, faint stars.  By comparing the pixel intensity histograms (PIHs)
of numerous artificial images constructed from model IMFs with the observed
PIH, we find that the best-fit model IMF for the Arches cluster has a cutoff mass
less than or similar to 0.1~M$_{\odot}$ and a shape very close to that of
the Kroupa MF.  Our findings imply that the IMF of the Arches cluster
is similar to those found in the Galactic disk.
\end{abstract}


\section{Introduction}

The Arches cluster is a young (2-4~Myr), compact ($\leq1$~pc), and
massive ($\sim2\times10^4$~M$_{\odot}$) star cluster located $26$~pc
away from the Galactic center (GC) in projection
\citep{fig99,kim00,fig02,naj04,mar08,esp09}. This extraordinary cluster has
been suspected to have had a shallower initial mass function (IMF) or an
elevated low-mass cutoff ($M_l$) in the IMF, compared to the IMFs inferred
for the Galactic disk clusters because of the
extreme environment of the GC, such as elevated temperatures and turbulent
velocities in the molecular clouds, strong magnetic fields, and large tidal
forces \citep{mor93}. The Quintuplet cluster is another young (3-5~Myr),
massive ($\sim10^4$~M$_{\odot}$) cluster in the GC area \citep{fig99}, but
it is much less dense in number density than the Arches,
and thus it is more difficult to
estimate its mass function from the photometry because of the confusion
problem with the foreground and background stars. The Arches cluster is an
excellent target for understanding the effects of the star-forming environment
on the IMFs of star clusters.

The first high-resolution photometry toward the Arches was performed
by \citet{fig99} using the infrared (IR) camera onboard the {\it Hubble
Space Telescope} (HST).  They estimated the cluster's mass function (MF)
down to $\sim 6$~M$_{\odot}$, which was the mass corresponding to the
50\% completeness limit of their photometry, $M_{50}$.  This has been
followed by several adaptive optics (AO) IR observations with a
few ground telescopes. \citet{sto02} obtained $M_{50}=5.6$~M$_{\odot}$
with the Gemini telescope, and \citet{sto05} acquired
$M_{50}=2.3$~M$_{\odot}$ with the Very Large Telescope (VLT). \citet{kim06}
obtained the lowest $M_{50}$ so far, 1.3~M$_{\odot}$, with the Keck
telescope. The IMF estimated in Kim et al. was only slightly shallower
(power-law exponent of $\Gamma = -1.0$ to $-1.1$) than the Salpeter IMF
(\citealt{sal55}; $\Gamma = -1.35$), and it did not indicate an
elevated low-mass cutoff, at least not down to 1.3~M$_{\odot}$.

Resolving stars fainter than 1.3~M$_{\odot}$ in the Arches cluster will
have to wait until the next-generation IR space telescope, such as
the {\it James Webb Space Telescope}, or an extremely large ground telescope,
such as the Giant Magellan Telescope, is available.  In the present
paper, we devise a method to estimate the shape of the low-end MF,
where the individual stars are not resolved, and apply it to the Arches
cluster.

In a conventional photometric method, stars in the observed image are
identified by fitting the point spread function (PSF).  However, our goal
here is to estimate the number distribution of stars fainter than
the photometric completeness limit (i.e., the stars that are too faint
to be fit with the PSF). Our new method involves
the histogram of pixel intensities in the observed image.

In most star clusters, faint stars are more abundant than bright ones. If the
number of unresolved stars is large enough, they will
have a non-negligible contribution to the pixel intensities and the number
distribution of pixel intensities in the observed image.  Thus, the pixel
intensity histogram (PIH) of an observed image can give valuable information
about unresolved, faint stars. With our method, we estimate the MF
of the Arches cluster below the photometric completeness limit by fitting
the PIHs of model images to the observed PIH.

We first build a synthetic luminosity function (LF) for the cluster by
combining the observed LF above the magnitude of the completeness limit
(limiting magnitude) and a model LF, which is converted from a model MF
below the limiting magnitude. We also build a synthetic LF for the
fore/background stellar population toward the Arches cluster using the
Milky Way star-count model of \citet{wai92}. We then create an artificial
image from these two synthetic LFs and compare the PIH of this image to
the observed PIH.  We try various model LFs below the limiting magnitude
until we find satisfactory agreement between the synthetic and observed PIHs.

Our new method does not directly yield an LF or MF of a cluster.
However, it can be used to reject certain MFs, and we find that
the IMF of the Arches cluster should not be too different from that
for the Galactic disk, such as the Kroupa IMF \citep{kro01}.

This paper is organized as follows.  In \S 2, we describe the observational
data and the photometric procedures used in the present study.  We discuss
the fore/background stellar population toward the Arches cluster and the
background flux in the observed images in \S 3.
\S 4 describes the procedures to search
for the most plausible MF of the Arches cluster. Summary and conclusion are
given in \S 5.

\section{Observational Data and Photometry}

\begin{figure}
\centering
\includegraphics[width=0.99\columnwidth]{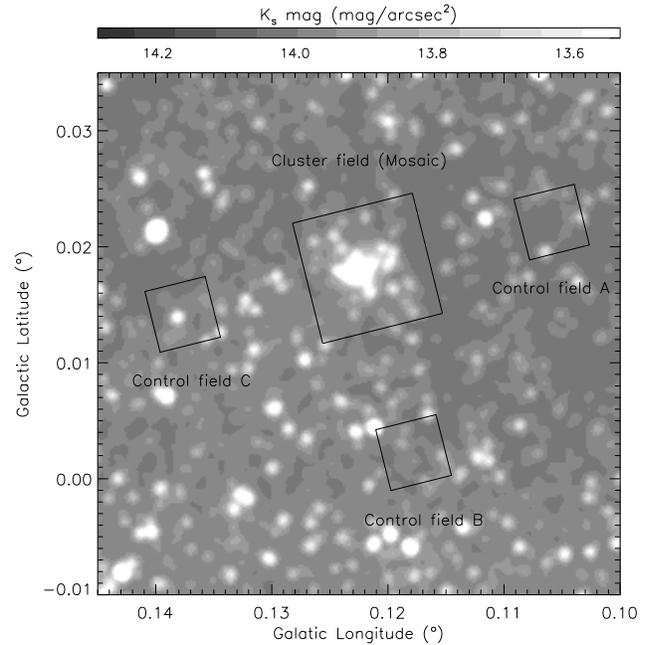}
\caption{Locations and image sizes of the cluster field and the three
control fields used in the present study.  All fields were observed by
the NICMOS2 camera onboard the {\it HST} \citep{fig99}. The cluster
field was observed in a $2\times2$ mosaic pattern.  The background,
a K$_{\rm s}$-band image from the 2MASS \citep{skr06}, is to show
that the three control fields have fairly uniform background emission
levels.}
\label{location}
\end{figure}

The imaging data of the Arches cluster that we use in this study were
obtained with the NICMOS2 camera onboard the {\it HST} on UT September
13/14, 1997 \citep{fig99}.  We chose this data set among others
not only because the PSFs of space observations are relatively stable
and well known, but also because this data set includes the images of
control fields near the target cluster.  The NICMOS2 camera has a
field-of-view (FOV) of $19''.2 \times 19''.2$ and $256 \times 256$ pixels
(each pixel covers $0''.076 \times 0''.076$ of the sky).  The Arches
cluster was observed in a mosaic-pattern of $2\times2$, and three control
fields were observed at a distance of $60''$ from the center of the cluster
field (Fig. \ref{location}).  All image data were taken with F110W, F160W, and
F205W filters, but the present study makes use of only F205W images
because they suffer the smallest extinction.  The exposure time of all
F205W images used in the present study is 256~s.

Data were reduced using STScI pipeline routines, and PSF photometry
(star-finding, PSF-building, and PSF-fitting) was performed using the DAOFIND
and DAOPHOT packages \citep{ste87} within the Image Reduction and Analysis
Facility (IRAF).\footnote{IRAF is distributed by the National Optical
Astronomy Observatories, which are operated by the Association of
Universities for Research in Astronomy, Inc., under cooperative agreement
with the National Science Foundation.}  The conversion of the PSF flux into
the Vega magnitude was done using the photometric keywords and zero-points
announced on the STScI
webpage\footnote{http://www.stsci.edu/hst/nicmos/performance/photometry}.

The photometry of faint stars is incomplete due to their low signals
and confusion by nearby brighter stars.  We estimated recovery fractions
for each 0.5 magnitude bin by planting artificial stars on the images.
The obtained LFs were corrected using the recovery fractions. In our
analyses in the forthcoming sections, only the portion of LF whose recovery
fraction is higher than 50\% is used.

We adopted a distance modulus of 14.52 for the GC ($R_g=8$~kpc; \citealt{rei93})
and solar-metallicity isochrone at 2 Myr of the Geneva model \citep{lej01} for
the conversion between the stellar magnitude and the mass. The reddening
value of each star was estimated from the color excess in F160W$-$F205W
color and the extinction law of \citet{rie89}. The average extinction
estimated for the cluster was 2.9~mag in F205W.

\subsection{The Control Fields}

\begin{figure*}
\begin{center}
\includegraphics[scale=1.0]{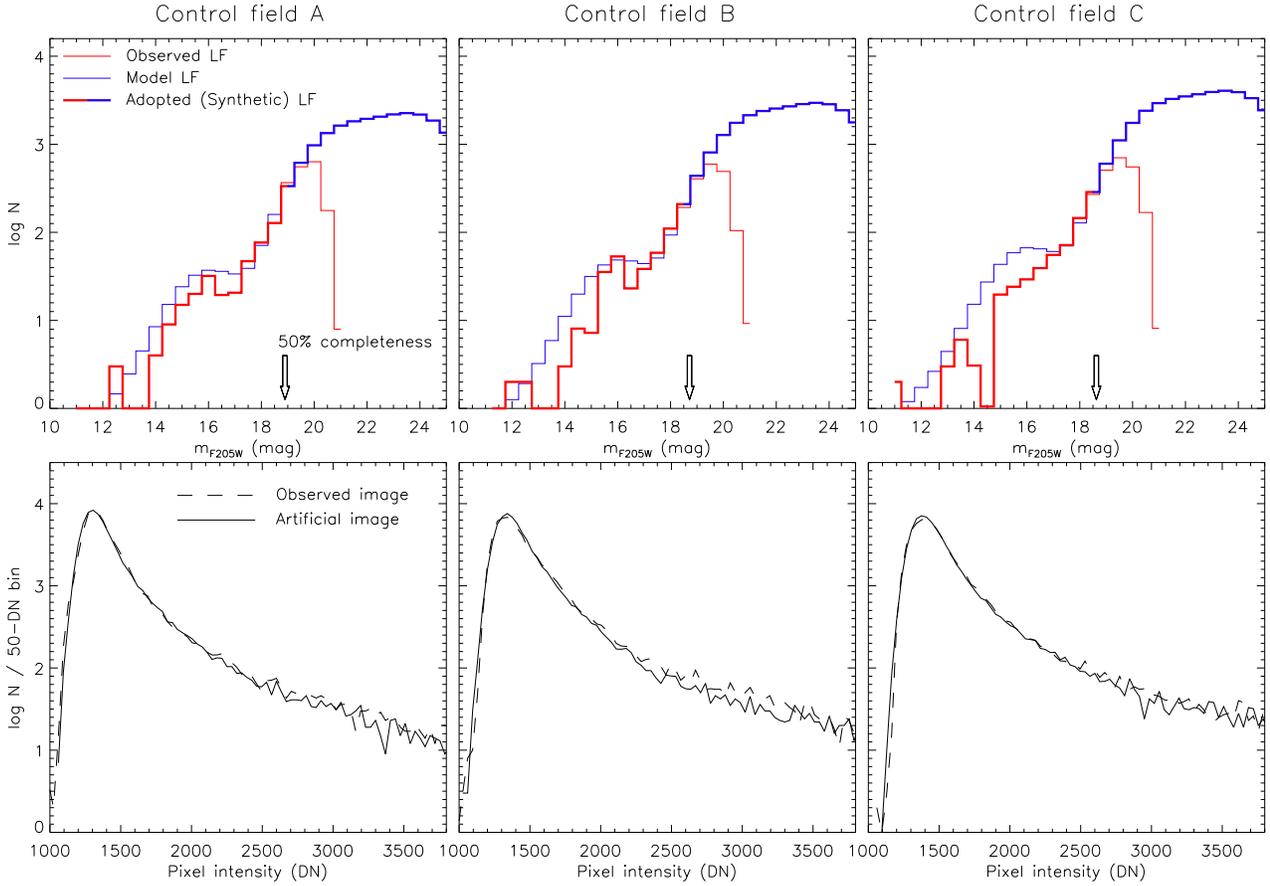}\
\caption{Luminosity functions (LFs) and pixel intensity histograms (PIHs)
of the control fields A, B, and C (left, middle, and right panels).
In the upper panels, the red lines are completeness-corrected
LFs from the observations, and the blue lines are the model LFs from the
star-count model of \citet{wai92}. We vertically shifted the model LFs
to match them with the observed LFs at $m_{50}$ (arrows), then built
the synthetic LFs (thick lines) by combining the observed LFs for the
bright-part and the model LFs for the faint-part. The lower panels compare
the observed PIHs with the PIHs of the artificial
images constructed from the corresponding synthetic LFs (solid lines).
The histograms are the number of pixels for each 50-DN bin.
\label{control_hist}}
\end{center}
\end{figure*}

Before constructing the fore/background LF toward the Arches
cluster, we first constructed the synthetic LFs in each direction of the three
control fields.  For the magnitude bins brighter than the magnitude of
$50\%$ completeness, $m_{50}$, of each field (18.9, 18.7, and 18.6 mag
for control fields A, B, and C, respectively), we adopted the
completeness-corrected F205W LFs from each field. For the magnitude
bins fainter than $m_{50}$, we constructed model LFs by integrating the model
number density of stellar objects within a given
apparent K-mag range along the pencil-beam volume element in the direction
of each control field.  For the model number density, we used the star-count
model of the Milky Way in the K band by \citet{wai92}, which
considers 87 different stellar types and various Galactic structures such
as the disk, bulge, halo, and spiral arms (here we assumed that the LFs in
the K and F205W bands are adequately similar).
The K magnitudes of stellar objects in the star-count model are
reddened with an average extinction of 2.9~mag and observed standard
deviations of 2.1, 1.6, and 1.8~mag for control fields A, B, and C,
respectively.

\begin{figure}
\begin{center}
\includegraphics[scale=1.0]{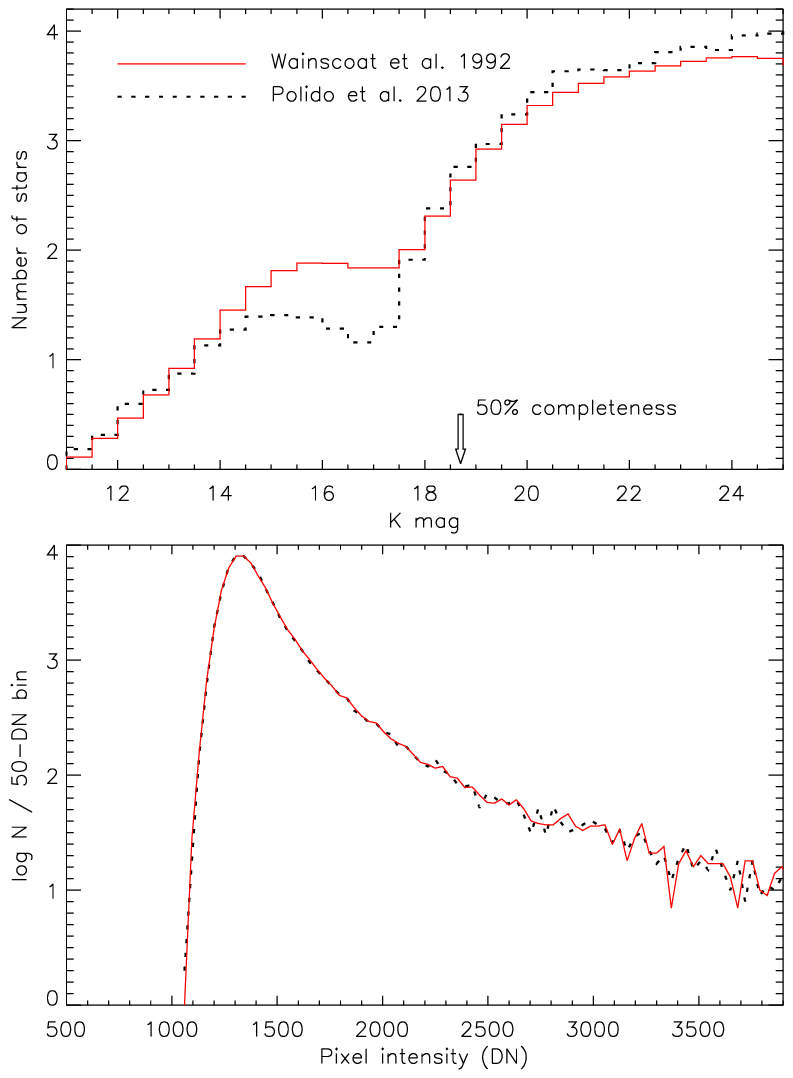}\
\caption{(Upper panel) Model LFs  toward the Arches cluster region by
\citet{wai92} and by \citet{pol13}. The arrow marks
the magnitude of 50\% completeness, $m_{50}$.  (Lower panel) PIHs
of the artificial images for the control field A that are constructed
with the star-count models by \citet{wai92} and by \citet{pol13}.
For the latter, the model log LF was vertically shifted downward
by 0.34 dex in order to match it with the observed LF at $m_{50}$.
\label{test}}
\end{center}
\end{figure}

\begin{figure}
\begin{center}
\includegraphics[scale=0.9]{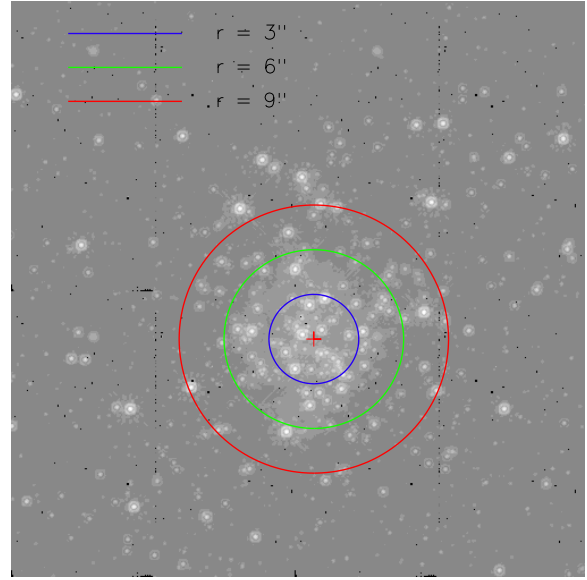}\
\caption{Cluster field image obtained with a F205W filter and the three annuli
of 3, 6, and $9''$ (corresponding to 0.12, 0.23, and 0.35~pc) from the
cluster center, which is marked with the cross.\label{annulus}}
\end{center}
\end{figure}

Ideally, the bright-part LF from the PSF photometry and the faint-part LF
from the star-count model should connect smoothly at $m_{50}$.  However,
the faint-part logarithmic LFs were found to be $\sim 0.25$~dex
higher than the bright-part log LFs at $m_{50}$.  This is thought to be
because the star-count model by \citet{wai92} is not accurate enough,
particularly toward the GC region, due to the large amounts of extinction.
For this reason, we subtracted 0.31, 0.28, and 0.19~dex from the
model log LFs of control fields A, B, and C, respectively, so the model LFs at
$m_{50}$ match with the observed LFs. Then we constructed the final
synthetic LF by merging the bright-part LF from the PSF photometry and the
modified faint-part LF from the star-count model.

The upper panels of Figure \ref{control_hist} show the completeness-corrected
LFs from observations (red lines), the model LFs modified (shifted vertically
as described above) to match the observed LFs at $m_{50}$ (blue lines),
and the final synthetic LFs for the three control fields (thick lines).

We then constructed the artificial images using these synthetic LFs for
each control field.  For this, we first built a set of 64 model PSFs at
$8\times8$ equally-spaced grid points of the detector plane using the
Tiny Tim,{\footnote{http://www.stsci.edu/hst/observatory/focus/TinyTim}}
a PSF modelling tool.  Each model PSF has $80\times80$ pixels.  Then
we planted artificial stars at random locations in the artificial images,
using whichever of the 64 model PSFs was closest  to the location of the
artificial star being planted.  The magnitude of the artificial star was
chosen randomly following the synthetic LF for each control field.
A dark current of $76.8~\mathrm{e}^-$ and readout noise with
a standard deviation of $\sqrt{26}~\mathrm{e}^-$ were added to the
artificial images.  Poisson uncertainties and a gain of
$5.4~\mathrm{e}^-/\mathrm{DN}$ [conversion between the electrons and the
data number (DN)] were considered when adding the artificial stars,
dark current, and background flux (discussed below).

The final artificial image was completed by adding a proper background
flux, which is rather difficult to estimate theoretically.  The
background flux includes telescopic thermal noise, zodiacal light,
external galactic sources, and cosmic IR background.  The instrument
handbook of the NICMOS camera predicts that thermal noise is the
dominant background source, with an estimated flux of $\sim 550$~DN
for an exposure of 256~s.  However, the flux level from the thermal
noise can vary over time, and the exact amount of background
flux from the zodiacal light and cosmic IR background in the F205W
band is not well known either.

Perhaps using the observed PIHs themselves is the best way to estimate
the background flux. The lower panels of Figure \ref{control_hist}
show the faint ends (DN smaller than 3,800) of the PIHs.  These faint-end
PIHs correspond to pixels
that do not belong to the main body of any bright, resolved stars
(the peak DN of a star with $m_{50}$ is $\sim3500$). The intensities
in this part of a PIH are mostly from various background sources and faint,
unresolved stars.  We found that adding a background flux of 1,190~DN
resulted in a good match between the PIHs from
observed and artificial images for all three control fields.

This demonstrates that we are able to reproduce the observed
PIHs in the directions of the three control fields with our procedures
for constructing the fore/background stellar LF and artificial images,
and that we are now ready to apply our procedures to the Arches cluster
to reveal the faint-part LF of the cluster itself.
The background flux of 1,190~DN obtained for the control fields
will be used for the cluster field in \S 4 as well.

The faint-end PIHs of our artificial images for the control
fields are statistically stable against the random choices of magnitudes
and positions for the artificial stars because the number
of pixels with $\mathrm{DN} \lsim 3,000$ in the artificial images
(and thus in the observed images as well) is large enough
($\sim50$-$\sim10,000$ for the 50-DN bin).

 To check if the PIHs of our artificial images are dependent on the
choice of the star-count model for the faint-part LF, we have also tried
the star-count model by \citet{pol13}, which is a modified version of
the model by \citet{ort93}.  The upper panel of Figure \ref{test} compares
the model LFs toward the Arches cluster region by \citet{wai92} and by
\citet{pol13}.  The shapes of the two LFs are quite similar at the
faint-part (below $m_{50}$).  Although the LF by Polido et al. is
slightly higher at the magnitudes below $m_{50}$, this does not matter
much because we vertically shift the model LF anyway to match it with
the observed LF at $m_{50}$.  The lower panel of Figure \ref{test}
compares the PIHs of the artificial images for the control field A
that are constructed with the two star-count models, and the two PIHs
are nearly indistinguishable.  This shows that our procedure of
constructing the artificial images for fore/background stars is
robust against the choice of star-count model for the faint-part LF.

\section{The Cluster Field}

\begin{figure*}
\begin{center}
\includegraphics[scale=1.0]{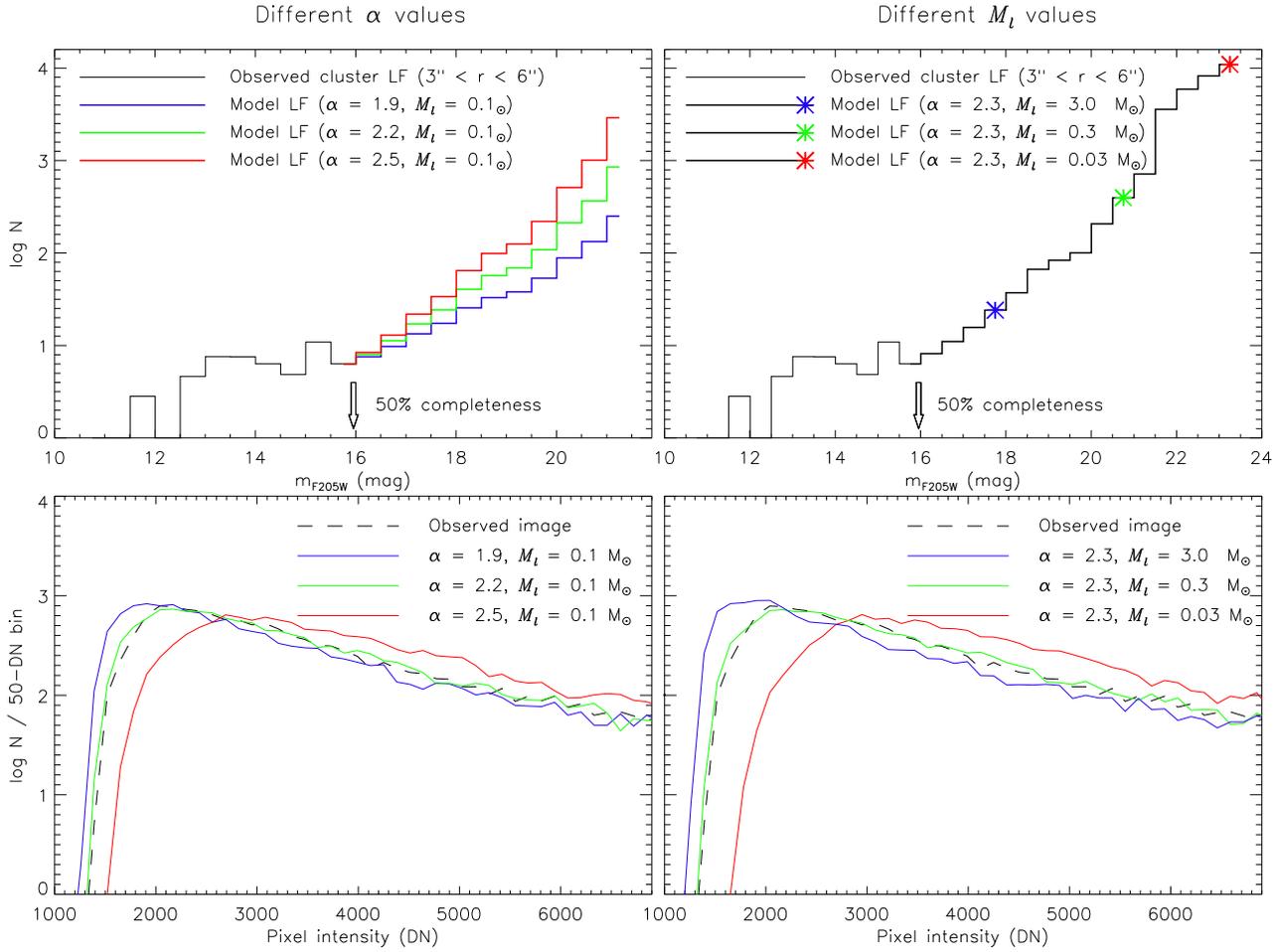}
\caption{LFs and PIHs of the cluster field with different $\alpha$
(left panels) or $M_l$ (right panels) values for single power-law MFs.
The model LFs, converted from the model MFs, are connected
with the observed cluster LF at $m_{50}$ (upper panels) to form the
synthetic LFs. The PIHs of the artificial images constructed from the
synthetic LFs are compared with the observed PIH (lower panels).
\label{single_diff}}
\end{center}
\end{figure*}

\begin{figure*}
\begin{center}
\includegraphics[scale=1.0]{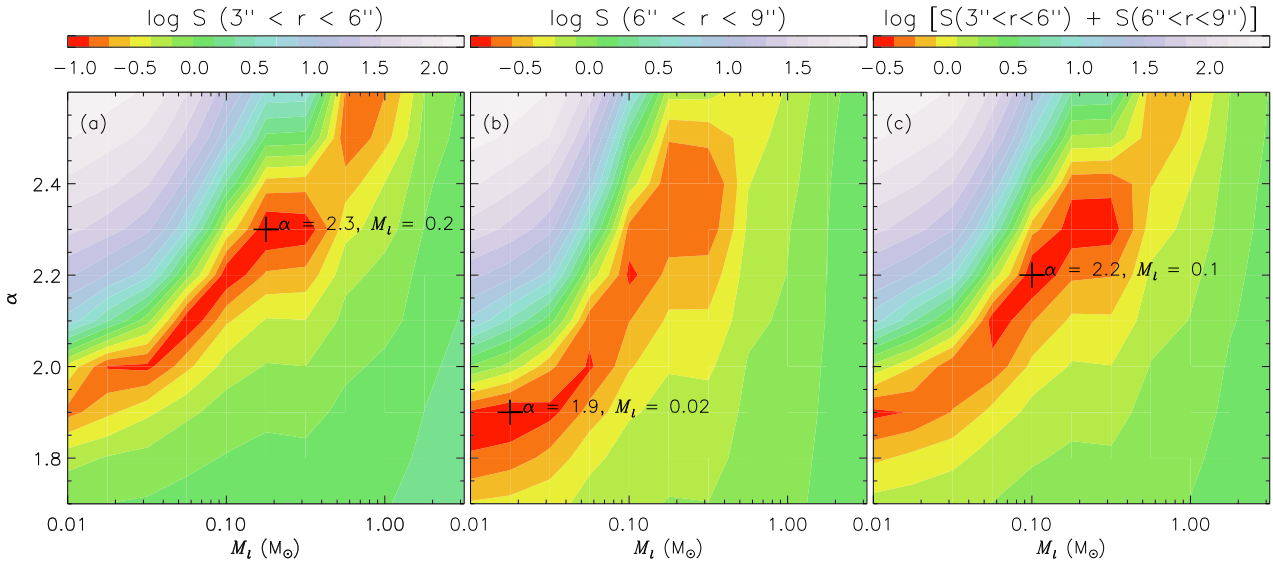}
\caption{Color maps of $\log S$ of the cluster field in $\alpha-M_l$ space
for a single power-law MF. The left and middle panels are for $3''<r<6''$
and $6''<r<9''$ annuli, respectively, and the right panel is for the
two annuli together.  The minimum $\log S$ points are marked with
cross symbols. \label{single_survey}}
\end{center}
\end{figure*}

Unlike in the control fields, in the cluster field the stellar number density
and LF are highly position-dependent. At the center of the cluster,
confusion by bright stars is so significant that the contribution of
the faint cluster stars to the PIH is almost negligible.
On the outskirt of the cluster, the photometry is limited by the
fore/background stars and the contribution of faint cluster stars
to the PIH becomes insignificant relative to the
fore/background stars.  Thus the best region for a study of faint cluster
stars is the intermediate region between the cluster center and the outskirt.
We chose two annuli ($3''<r<6''$ and $6''<r<9''$) for our analyses here.
[Fig. \ref{annulus}; we adopted the position of the cluster center used in
\citet{fig99}; note that \citet{kim00} estimates the tidal radius of the
cluster to be $\sim 26''$].

Our stellar photometry and completeness test resulted in $m_{50}=16.0$~mag
for $3''<r<6''$ and 17.4~mag for $6''<r<9''$.  As in the control fields,
our synthetic LFs for the two cluster annuli are composed of two parts:
the bright-part LF from our PSF photometry for magnitude bins brighter
than $m_{50}$ and the faint-part LF from a model for magnitude
bins fainter than $m_{50}$.

The faint-part model LF consists of 1) the model LF for the
fore/background stars and 2) the LF converted from a model MF for the
cluster stars.  For the fore/background model LF, we used the average
of the three synthetic LFs for the control fields.  For the cluster
model LF, we tried the following two models: 1) LFs converted from single
power-law MFs with various exponents and lower mass limits, and
2) LFs converted from the Kroupa MF \citep{kro01} with various lower
mass limits.

We constructed artificial images for the two cluster annuli following
the same procedure used for the control fields.  However, the area covered
by each cluster annulus is rather small, and we found that the shape of
the faint-end PIH depended slightly on the random choices of magnitudes
and positions of bright artificial stars.  Therefore, we used
the magnitudes and positions of the actual, observed stars when planting
artificial stars brighter than $m_{50}$ instead of randomly selecting their
magnitudes and positions from the observed LF.

A background flux of 1,190~DN was applied to the artificial images
for the cluster annuli, which is the value deduced from the control
field images.

\subsection{Single Power-Law Mass Function}

In this subsection, we describe our attempts with the single power-law MFs
with various power-law exponents $\alpha$ ($dN\propto~M^{-\alpha}dM$)
and $M_l$ for the faint-part LF of the cluster.
The upper panels of Figure \ref{single_diff} show some sample synthetic
LFs for the $3''<r<6''$ annulus, whose faint-part cluster LFs are from the
MFs with three different $\alpha$ values (left panel), and three
different $M_l$ values (right panel).

The lower panels of Figure \ref{single_diff} plot the faint-end PIHs
of the observed image (dotted lines) and those of the artificial images
constructed from the synthetic LFs in the corresponding
upper panels (solid lines).  First, the heights of PIHs from the artificial
images are generally consistent with those from the observed image, implying
that our procedure of constructing artificial images for the cluster
field has no fundamental flaws.  Second, the turnover intensity of
the PIH clearly depends on the $\alpha$ and $M_l$ values (though it
is less sensitive on small $\alpha$ and large $M_l$ values).
This second point demonstrates that the PIHs can be used to reject or accept
assumed faint-end LFs (and thus MFs) for the Arches cluster.

The few sample experiments shown in Figure \ref{single_diff} already imply
that the best-fit $\alpha$ and $M_l$ are around 2.2-2.3 and
0.1-0.3~M$_\odot$, respectively.  To more quantitatively search
for the $\alpha$ and $M_l$ values that best fit the observed PIH, we devised a
statistic:
\begin{equation}
	S = \left < \left(\log N_{\mathrm obs}(I)-\log N_{\mathrm art}(I)
                    \right)^2 \right >_{N>100},
\end{equation}
where $N_{\mathrm obs}(I)$ and $N_{\mathrm art}(I)$ are the numbers of
pixels with an intensity (DN) $I$ in the observed and artificial images,
respectively.  The bracket indicates an average, and we chose to
average only numbers with $N(I)>100$.  This rather unusual
statistic was adopted because the logarithmic shape of the PIH better
discerns different $\alpha$ and $M_l$ values than the linear shape does.

The color map of $\log S$ in the $\alpha$-$M_l$ space for the two
cluster annuli are shown in panels a and b of Figure 5.  The minimum
$\log S$ is near ($\alpha=2.3$, $M_l=0.2 \msun$) for the inner
annulus and near ($\alpha=1.9$, $M_l=0.02 \msun$) for the outer annulus.
The $\log S$ contours near the minima are rather elongated along the
curves connecting the (large $\alpha$, large $M_l$) and (small $\alpha$,
small $M_l$) corners because those curves have similar contributions
(roughly in numbers) of faint stars to the faint-ends of the PIHs.

The simplest way to estimate the best-fit $\alpha$ and $M_l$ values
from the two annuli simultaneously is perhaps by adding the
$S$ values from the two annuli.  The sum of the two $S$ values
has a minimum at ($\alpha=2.2$, $M_l=0.1 \msun$) and has less
elongated contours around the minimum (Fig. \ref{single_survey}c).

Our simple analyses with single power-law MFs imply that the present-day
MF of the Arches is not significantly different from the Salpeter MF
whether the two annuli are considered separately or simultaneously.

\begin{figure}
\begin{center}
\includegraphics[scale=1.0]{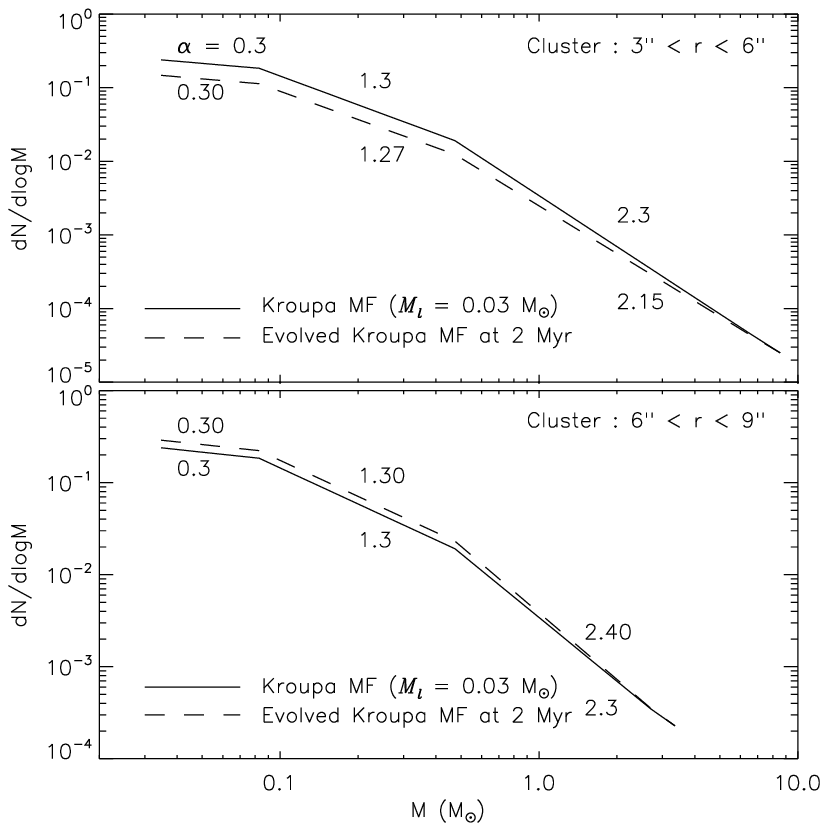}
\caption{Original Kroupa MF (solid lines) and the dynamically evolved Kroupa
MF at 2~Myr calculated using the Fokker-Planck model (dashed lines)
below the masses that correspond to $M_{50}$ for annuli of $3''<r<6''$
(upper panel) and $6''<r<9''$ (lower panel).  The heights of the two MFs
are adjusted to match at $M_{50}$. \label{kroupa_mf}}
\end{center}
\end{figure}

\begin{figure*}
\begin{center}
\includegraphics[scale=1.0]{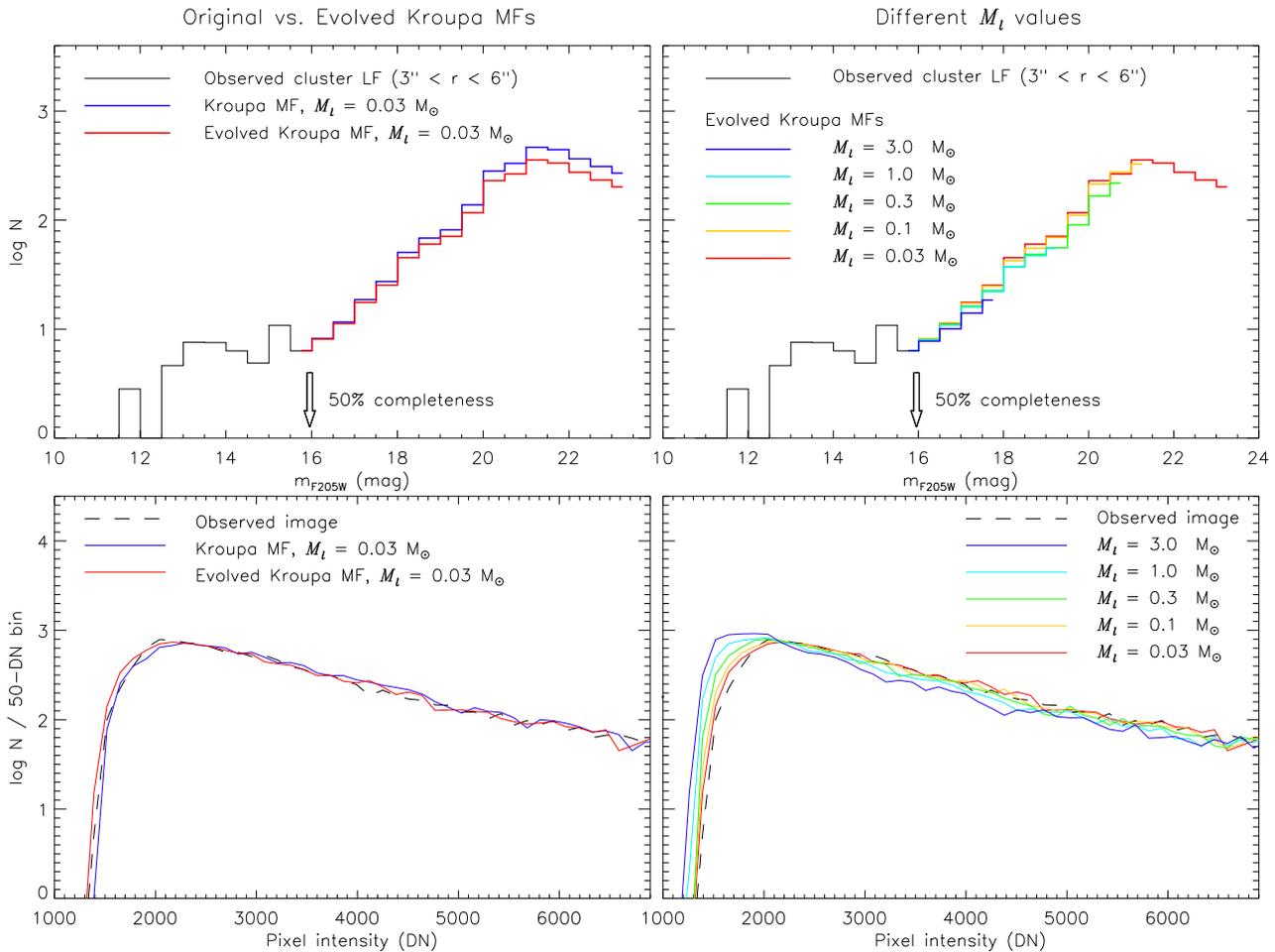}
\caption{LFs and PIHs of the cluster field ($3''<r<6''$) for the original
Kroupa MF and the evolved Kroupa MF at 2~Myr (left panels) and for the
five different $M_l$ values for the evolved Kroupa MF (right panels).
\label{kroupa_diff}}
\end{center}
\end{figure*}

\begin{figure}
\begin{center}
\includegraphics[scale=1.0]{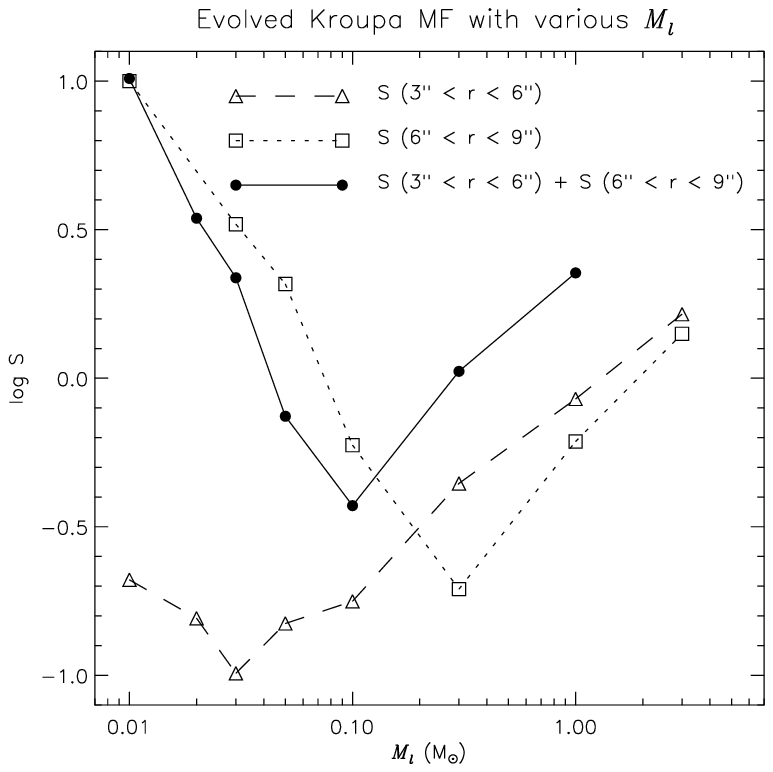}
\caption{$\log S$ values as functions of $M_l$ for the evolved
Kroupa MFs at 2~Myr. Triangles and squares are for the inner ($3''<r<6''$)
and outer ($6''<r<9''$) annuli, respectively.  Filled circles are for
the two annuli together.\label{kroupa_survey}}
\end{center}
\end{figure}

\subsection{Kroupa Mass Function}

In this subsection, we use more realistic MF models for the faint-part
LF of the cluster.  We adopted the Kroupa MFs \citep{kro01} with various
$M_l$ as the initial MF of the cluster, and allowed them to dynamically evolve
for 2~Myr with the anisotropic Fokker-Plank (F-P) model of \citet{kim99}
and \citet{kim00}.  The Kroupa MF consists of three
power-laws, $\alpha=2.3$ for $M>0.5$~M$_{\odot}$, $\alpha=1.3$ for
$0.08<M/\msun<0.5$, and $\alpha=0.3$ for $M<0.08$~M$_{\odot}$.
Following \citet{kim99}, we adopted the following for the initial
conditions of the F-P calculation: density and velocity dispersion
structures from the King model, total mass of $2\times10^4$~M$_{\odot}$,
upper mass boundary of 150~M$_{\odot}$, tidal radius of 1.1~pc, and the
King concentration parameter of 4.

Figure \ref{kroupa_mf} shows the Kroupa IMF with $M_l=0.03$~M$_{\odot}$
and its evolved MF at 2~Myr from our F-P calculation for the two cluster
annuli.  The MF at 2~Myr is slightly steeper in the inner annulus, while
it is slightly shallower in the outer annulus due to the mass segregation.

The synthetic LFs are constructed in the same way as in \S 4.1: the
observed LF is adopted for the bright-part, and the faint-part LF is
converted from the model MF and joined to the bright-part LF at $m_{50}$.
The artificial images are then created in the manner described in \S 4.1.

The synthetic LFs and PIHs from the artificial images for the inner
cluster annulus are shown in the left panels of Figure \ref{kroupa_diff}
for the two model MFs: the Kroupa IMF and its evolved MF.  The difference
in PIHs between the two model MFs are insignificant because the evolution
of the MF is not considerable for the first 2~Myr in the radial ranges of our
consideration, as seen in Figure \ref{kroupa_mf}.

The right panels of Figure \ref{kroupa_diff} compare the LFs and PIHs for the
five evolved MFs with $M_l$ values between $0.03 \msun$ and $3 \msun$.
The faint-part slopes of the LFs differ slightly for the different $M_l$ values
because the degree of mass segregation depends on $M_l$.  Among the five
PIHs, the one for $M_l=0.03 \msun$ best fits the observed PIH.  MF models
with $M_l \gsim 0.3\msun$ considerably underestimate the PIH.

Figure \ref{kroupa_survey} plots the profiles of the $S$ values as a function of
$M_l$ for the inner and outer annuli as well as the summed $S$ profile for the
two annuli.  The minimum $S$ values are found at $0.03 \msun$ for the inner
annulus and $0.3 \msun$ for the outer annulus.  When the two annuli
are considered simultaneously, the minimum $S$ is found at $0.1 \msun$.
This clearly shows that a significant number of stars below $1 \msun$
are required to match the observed PIH of the Arches cluster.

\section{Summary}

In the present paper, we described a novel photometric method that compares
PIHs between an observed image and artificial images constructed with
various synthetic LFs to estimate the shape of the low-end MF
for the Arches cluster. The synthetic LFs were constructed by combining
the observed LF above $m_{50}$ and a model LF below $m_{50}$, which was
converted from two different types of model MFs: single power-law MFs with
various $\alpha$ and $M_l$ and the Kroupa MFs with various $M_l$.

In the case of the single power-law MF, we found the best-fits between the
observed and artificial PIHs at ($\alpha=2.3$, $M_l=0.2\msun$) for the
inner annulus of $3''<r<6''$ and at ($\alpha=1.9$, $M_l=0.02\msun$) for
the outer annulus of $6''<r<9''$.  When the inner and outer
annuli are considered together, the best-fit $\alpha$
and $M_l$ values are 2.2 and $0.1\msun$, respectively, which is very
close to the Salpeter IMF.

As a more realistic IMF model, we adopted the Kroupa IMF \citep{kro01}
and used its dynamically evolved MFs calculated with the Fokker-Planck
method.  The best-fits between the observed and artificial PIHs were found at
$M_l=0.03\msun$ for the inner annulus, $M_l=0.3\msun$ for the outer annulus,
and $M_l=0.1\msun$ for both annuli.  This shows that the Arches cluster
has a significant number of stars that are not resolved by current
IR telescopes and that the most likely IMF of the Arches cluster is
not too different from the IMF found in the Galactic disk.

\section*{Acknowledgments}

We thank Professor Myong Gyoon Lee for helpful discussions and comments.
J.S. appreciates Seungkyung Oh for the help with the IRAF package. S.S.K's
work was supported by the Mid-career Research Program (No. 2011-0016898)
through the National Research Foundation (NRF) grant funded by the
Ministry of Education, Science and Technology (MEST) of Korea.

\end{document}